\documentstyle[prd,aps,epsfig,psfig]{revtex}
\tighten
\begin{document}
\draft
\twocolumn[\hsize\textwidth\columnwidth\hsize\csname
@twocolumnfalse\endcsname
\title{Topological lens effects in Universes with Non-Euclidean Compact
Spatial Sections}
\author{Roland Lehoucq$^1$, Jean-Pierre Luminet$^2$ and 
Jean-Philippe Uzan$^{3,2}$}
\address{(1) CE-Saclay, DSM/DAPNIA/Service d'Astrophysique,\\
 F-91191 Gif sur Yvette cedex (France)\\
(2) D\'epartement d'Astrophysique Relativiste et de Cosmologie,\\
Observatoire de Paris, UPR 176, CNRS, F-92195 Meudon (France).\\
(3) D\'epartement de Physique Th\'eorique, Universit\'e de Gen\`eve,\\
24 quai E. Ansermet, CH-1211 Geneva (Switzerland)}
\date{\today}
\maketitle
\begin{abstract} 
Universe models with compact spatial sections smaller than the
observable universe produce a topological lens effect. Given a catalog
of cosmic sources, we estimate the number of topological images in
locally hyperbolic and locally elliptic spaces, as a function of the
cosmological parameters, of the volume of the spatial sections and of
the catalog depth.  Next we apply the crystallographic method, aimed
to detect a topological signal in the 3D distance histogram between
images, to compact hyperbolic models. Numerical calculations in the
Weeks manifold allows us to check the absence of crystallographic
signature of topology, due to the fact that the number of copies of
the fundamental domain in the observable covering space is low and
that the points are not moved the same distance by the holonomies of
space.
\end{abstract}
\pacs{Key words~: cosmology: large scale structure of the universe; 
topology.}
\vskip2pc]

\section{Introduction}\label{INTRO}

The question of whether our universe has a finite spatial extension or
not is still an open question related to the topology of the universe
(see \cite{luminet95} for a review and \cite{cqg98} for latest
developpements). Recently, there has been a large activity to
constrain and/or to observe the shape and the size of the
universe. Many methods have been proposed to detect its spatial
topology using catalogs of discrete sources (clusters of galaxies
\cite{lehoucq96}, quasars\cite{roukema96}) and the cosmic microwave
background\cite{stevens93,oliveira96,cornish97,uzan98}.  All the
methods rest on a ``topological lens" effect which generates multiple
images of cosmic sources, as soon as the compact spatial sections have
a volume smaller than the observable universe.  In the past, the idea
of using the topological images was extensively applied to universe
models with Euclidean (see
e.g. \cite{fang83,demianski87,fagundes87,ellis71}) and hyperbolic (see
e.g. \cite{gott80}) spatial sections . More recently, the crystallographic
method \cite{lehoucq96}, which relies on the existence of topological
images whatever the underlying geometry, was applied only to locally
flat universes, and was able to put a bound on the characteristic size
$L$ of Euclidean space to $L\leq 650\,h^{-1}\hbox{Mpc}$ (with
$h=H_0/100\,\hbox{km/s/Mpc}$, $H_0$ being the Hubble constant).  The
efficiency of the crystallographic method obviously depends on the
number of topological images of a given object within the horizon size
or within the limits of current catalogs used for the test.  The
applicability of the method in Euclidean space has also been discussed
by Fagundes and Gausmann \cite{fagundes97} when the size of the
physical space is comparable to the horizon size.

We can naturally wonder if the method applies as well in locally
hyperbolic or elliptic manifolds, and if we can get any constraint on
the size of space from the existing catalogs of cosmic objects.
We keep also in mind the growing weight of observational evidence for
a low density universe (see e.g. Spergel in \cite{cqg98}).  Thus in
this article we focus  mainly on universes with locally
hyperbolic compact spatial sections \cite{thurston}.  The universe is
described by a 4-manifold ${\cal M}$ and a Lorentzian metric ${\bf g}$
and we assume that ${\cal M}$ can be splitted as ${\cal
M}=\Sigma\times R$ (see e.g. \cite{hawking73} for the conditions of
such a splitting). As any  multi-connected closed
three-dimensional manifold, $\Sigma$ can be described by its
fundamental domain (a polyhedron) and its holonomy group $\Gamma$,
which identifies the faces of the polyhedron by pairs
\cite{luminet95}. Such hyperbolic manifolds $\Sigma=H^3/\Gamma$ have a
remarkable property that links topology and geometry~: the {\it
rigidity theorem} \cite{mostow73,prasad73} implies that geometrical
quantities such as the volume, the lengths of its closed geodesics,
\ldots, are topological invariants. The volume of the manifold can
then be used to classify these manifolds \cite{thurston}. The volumes
of compact hyperbolic manifolds are bounded below \cite{gabai96} by:
\begin{equation}
\hbox{Vol}(\Sigma)\geq \hbox{Vol}_{\rm min}\simeq0.166,
\end{equation}
in units of the curvature radius.

The smallest known compact hyperbolic manifold, likely 
to produce the greatest topological lens effects, is the Weeks
space \cite{weeks85,matveev} (see Appendix A for a description), such that
\begin{equation}
\hbox{Vol}=0.94272,\quad r_+=0.7525,\quad r_-=0.5192,
\end{equation}
where $r_+$ and $r_-$ are respectively the radii
of the largest (smallest) geodesic ball that contains (is contained in)
the fundamental domain.

The fundamental domain and the holonomy group of the known
three--dimensional compact hyperbolic manifolds can be found by using
the software {\it SnapPea} \cite{weeks} which gives all the
informations needed to compute the topological lens effects, such as
the volume, the generators of the holonomy group, the lengths of
closed geodesics.

Fagundes \cite{fagundes93} already used a universe whose spatial
sections had the topology of the Weeks manifold to discuss the
controversy about the quasars redshifts. In his paper, he gave an
interesting description of the fundamental domain and of the holonomy
group. The same author previously studied a 2+1 hyperbolic cosmology
\cite{fagundes85a} and a universe with hyperbolic spatial sections
whose fundamental domain was a hyperbolic icosahedron
\cite{fagundes89} (known as the Best space \cite{best71}) to
investigate the same problem.

In section \ref{CRYST} we describe the crystallographic method and the
way to implement it, focusing on the interface with {\it SnapPea}.  In
\S \ref{APPL}, we discuss the applicability of this method in
compact hyperbolic universes (\S \ref{CHM}) and in compact elliptic
universes (\S \ref{CEM}) and specially the dependence of the number of
images in function of the density parameter and of the cosmological
constant.  We then give (\S \ref{NUM}) the numerical results of simulations in the 
Weeks manifold, and discuss the influence of the catalog, of the density
parameter and of the position of the observer within the fundamental 
domain.

\section{The crystallographic method in compact hyperbolic 
universes}\label{CRYST}

In this section, we describe the crystallographic method and the way
to implement it to universes with compact hyperbolic spatial sections.
The local geometry of such a universe is described by a
Friedmann-Lema\^\i tre metric
\begin{equation}
ds^2=-dt^2+a^2(t)\left(d\chi^2+\sinh^2{\chi}d\Omega^2 \right).\label{METR}
\end{equation}
where $a$ is the scale factor, $t$ the cosmic time and
$d\Omega^2\equiv d\theta^2+\sin^2{\theta}d\varphi^2$ the infinitesimal
solid angle.

A locally hyperbolic three--dimensional manifold can be embedded in
four--dimensional Minkowski space by introducing the
set of coordinates $(x^\mu)_{\mu=0..3}$ related to the
intrinsic coordinates $(\chi,\theta,\varphi)$ through
(see e.g. \cite{wolf67,coxeter65}) 
\begin{eqnarray}
x_0&=&\cosh{\chi}\nonumber\\
x_1&=&\sinh{\chi}\sin{\theta}\sin{\varphi}\nonumber\\
x_2&=&\sinh{\chi}\sin{\theta}\cos{\varphi} \nonumber\\
x_3&=&\sinh{\chi}\cos{\theta},
\end{eqnarray}
so that the three--dimensional hyperboloid $H^3$ 
has the equation
\begin{equation}
-x_0^2+x_1^2+x_2^2+x_3^2=-1.
\end{equation}
[Note that when $a(t)=t$, the line element (\ref{METR}) describes a Milne 
universe which, using the coordinates transformation $(t'=t\cosh\chi, 
r=t\sinh\chi)$ reduces to the Minkowski line element in spherical coordinates.  
This can describe a $\Omega=0$ open cosmology (see e.g. \cite{misner73}).]

With these notations, the comoving spatial distance between two points
of comoving coordinates $x$ and $y$ can be computed directly in the
Minkowski space by
\cite{fagundes89}
\begin{equation}
d[x,y]=\arg\cosh{\left[\frac{x^\mu y_\mu}{\left(x^\mu x_\mu \right)^{1/2}
\left( y^\mu y_\mu \right)^{1/2}}\right]},
\end{equation}
where $x_\mu=\eta_{\mu\nu}x^\nu$, $\eta_{\mu\nu}$ being the
Minkowskian metric. Note that Minkowski space can be
mapped onto the interior of an ordinary sphere $S^2$ of unit radius by using the
Klein coordinates $(X_i)_{i=1..3}$ \cite{wolf67,coxeter65} defined by
\begin{equation}
X_i=x_i/x_0.\label{Kleincoord}
\end{equation}

The universal covering space being described, we now choose a
topology, i.e. a holonomy group $\Gamma$ such that the spatial
sections are
${\Sigma}=H^3/\Gamma$. ${\Sigma}$ can be described by its fundamental
domain whose $2K$ faces are identified by pairs by the elements of
$\Gamma$. $\Gamma$ has $2K$ generators which, in the case of the Weeks
manifold ($K=9$) can be obtained from {\it SnapPea} and are given
in appendix A.

Indeed, the elements of $\Gamma$ are isometries so that
\begin{eqnarray}
\forall x,y\in\Sigma\quad \forall g\in\Gamma,\quad
\hbox{dist}[x,y]=\hbox{dist}[g(x),g(y)].
\end{eqnarray}

The crystallographic method \cite{lehoucq96} is based on a property
of multi--connected  universes according to which each topological image of a given object
is linked to each other one by the holonomies of space. Indeed, we do
not know these holomies as far as we have not determined the topology,
but we know that they are isometries. For instance in locally Euclidean universes, 
to each holonomy is associated a distance $\lambda$, equal to the 
length of the translation by which the fundamental domain is 
moved to produce the tessellation in the covering space. Assume the 
fundamental domain contains $N$ objects (e.g.  galaxy clusters), 
if we calculate the mutual 3D--distances between every pair of topological images (inside the particle 
horizon), the distances $\lambda$ will occur $N$ times for each copy of the 
fundamental domain, and all other distances will be spread in a smooth way between 
zero and two times the horizon distance.  In a histogram
 plotting the number of pairs versus their 3D separations, 
the distances $\lambda$ will thus produce peaks. Simulations indeed 
showed that the pairs between two topological images of the same 
object drastically emerge from ordinary pairs\cite{lehoucq96} in the histogram.

Two kind of catalogs of astronomical objects can be thought of to
apply this method : the galaxy cluster catalogs, which typically have
a redshift depth $z=1$, and the quasars catalogs, which typically
extend to $z=3$. Concerning quasars, even if their lifetime is
probably too short to be good candidates for producing topological
images, they are usually part of systems that have a much larger
lifetime \cite{paal}. The angular resolution needed is given by the
fact that the objects have a peculiar velocity and that they will not
be seen at exactly the same position \cite{luminet95}. Note that the
crystallographic method, contrary to the ``direct'' method which would
try to recognize topological images of individual objects, is not
plagued by the evolution problem, i.e. that topological images of the
same object are seen at different stages of its evolution.

\section{Estimation of the number of topological images in 
non-Euclidean universes}
\label{APPL}

\subsection{Compact hyperbolic universes}\label{CHM}

To estimate the applicability of the crystallographic method in
hyperbolic compact universes, we estimate the number of topological
images of a given object up to a redshift $z$. With the metric (\ref{METR}),
the Einstein equations reduce to the Friedmann equation
\begin{equation}
H^2=\kappa\frac{\rho_m}{3}-\frac{K}{a^2}+\frac{\Lambda}{3},\label{fried1}
\end{equation}
$\rho_m$ being the matter density, $\Lambda$ the cosmological constant
and $\kappa\equiv 8\pi G/c^4$.  $H$ is the Hubble constant defined by
$H\equiv \dot a/a$ with $\dot X\equiv \partial_tX$. We choose the
units such that the curvature index is $K=-1$. Introducing
$\Omega_\Lambda\equiv \Lambda/3H^2$, $\Omega_m\equiv\kappa\rho_m/3H^2$
and the redshift $z$ defined by $1+z\equiv a_0/a$, (\ref{fried1}) can
be rewritten as (see e.g. \cite{peebles93})
\begin{equation}
\frac{H^2}{H_0^2}=\Omega_{m0}(1+z)^3+\Omega_{\Lambda0}+(1-\Omega_{m0}-
\Omega_{\Lambda0})(1+z)^2.\label{fried2}
\end{equation}
For that purpose we have used equation (\ref{fried1}) evaluated today (i.e.
at $t=t_0$) and we have assumed that we were in a matter dominated universe
so that $\rho_m\propto a^{-3}$ [this hypothesis is very good since
we restrict ourselves to small redshift].

The radius of the observable region at a redshift $z$ is given by
integration of the radial null geodesic equation $d\chi=dt/a$ and
reads
\begin{eqnarray}
\chi(z)&\equiv&\int_{a_0}^a\frac{da}{a\dot a}\nonumber\\
&=&
\int_{\frac{1}{1+z}}^1\frac{\sqrt{1-\Omega_{m0}-\Omega_{\Lambda0}}dx}
{x\sqrt{\Omega_{\Lambda0}x^2+(1-\Omega_{m0}-\Omega_{\Lambda0})
+\frac{\Omega_{m0}}{x}}}.\label{chi1}
\end{eqnarray}
(\ref{chi1}) is integrated numerically and the result can be compared,
when $\Omega_\Lambda=0$, to the analytic expression
(see e.g. \cite{gradstheyn})
\begin{equation}
\chi(z)=\left[\arg\cosh{\left(1+\frac{2(1-\Omega_{m0})}{\Omega_{m0}}x
\right)}\right]^1_{\frac{1}{1+z}}.\label{chi_de_z}
\end{equation}
The number of
topological images of a given object at a redshift $z$ can be
estimated by computing the ratio between the volume of the geodesic
sphere of radius $\chi(z)$ and the volume of the manifold which
is a topological invariant.  This leads to
\begin{equation}
N(\Omega_{m0},\Omega_{\Lambda0};z<Z)=\frac{\pi\left(\sinh{2\chi(Z)}
-2\chi(Z)\right)}{\hbox{Vol}(\Sigma)}.
\end{equation}
It can be easily understood that this under-estimates the number
of images.\\

\begin{figure}
\centering
\epsfig{figure=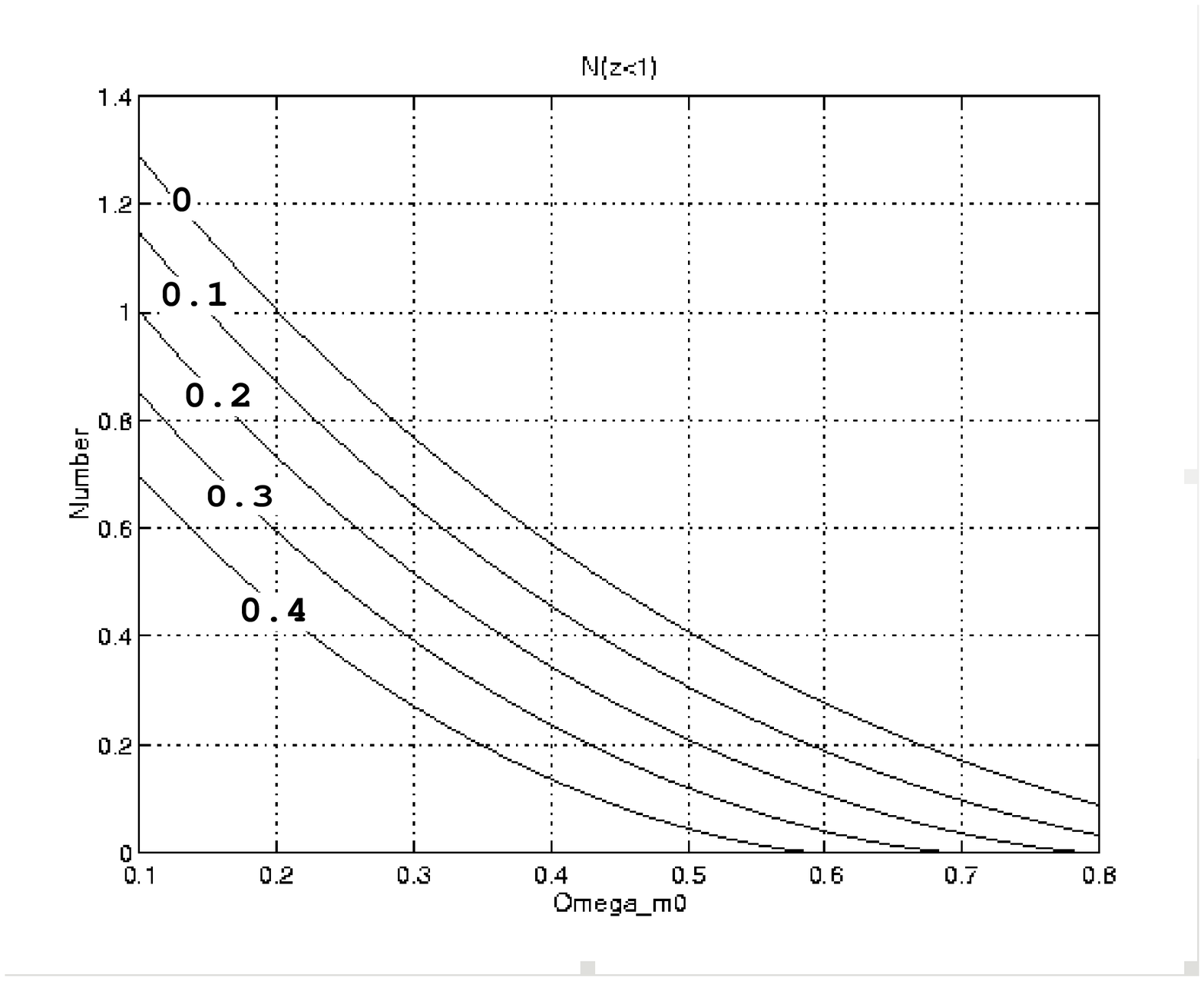, width=8cm}
\caption{Number of topological images in a catalog of clusters
up to a redshift $z=1$ in function of $\Omega_{m0}$
and $\Omega_{\Lambda0}$ in a universe whose spatial sections have
the topology of the Weeks manifold. The curves are labeled by the 
values of $\Omega_{\Lambda0}$. Topology is detectable only if the 
number of topological images is greater than unity.}
\label{plotz1}
\end{figure}

\begin{figure}
\centering
\epsfig{figure=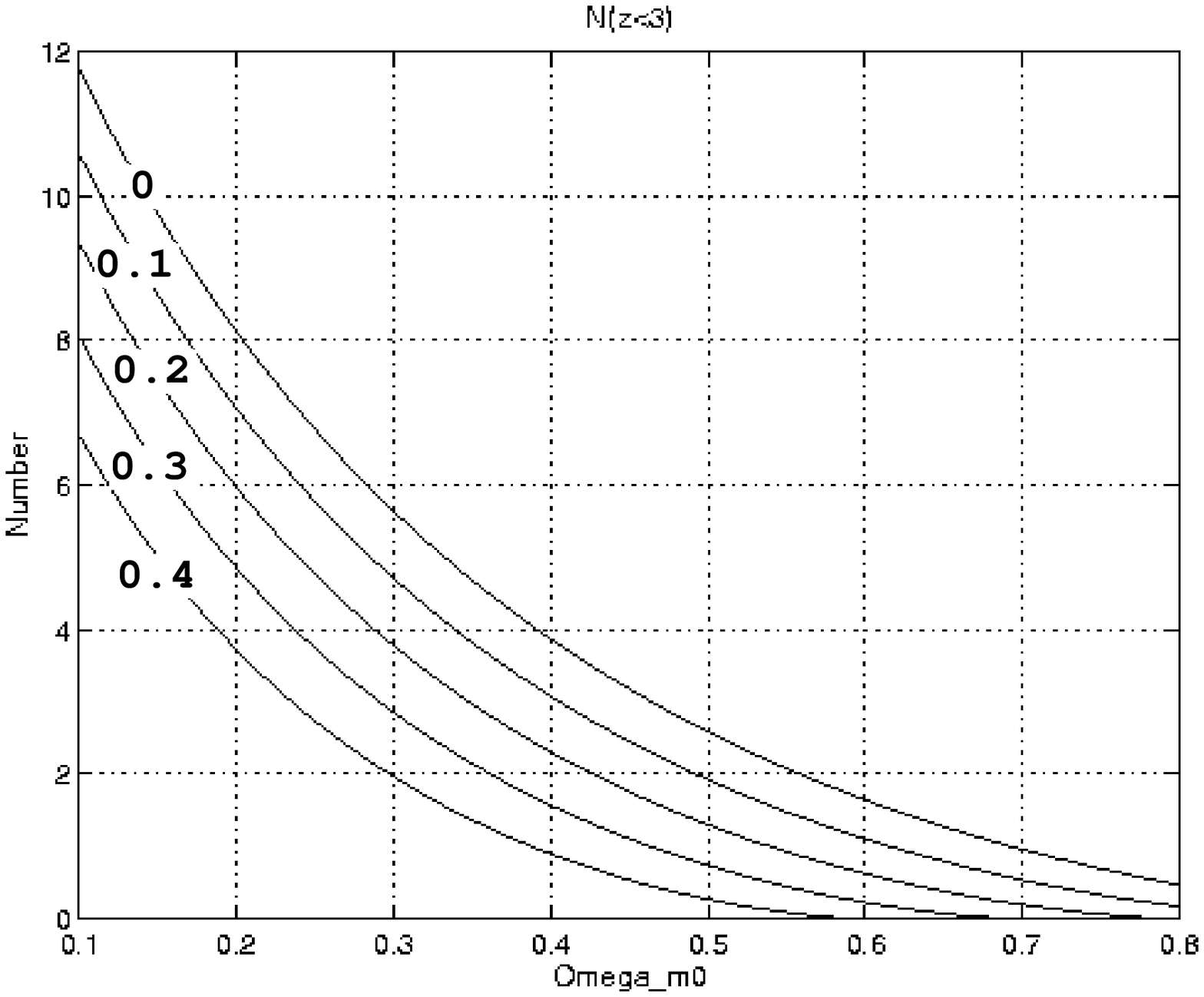, width=8cm}
\caption{The same as Figure 1 in the case of a catalog of quasars 
extending up to a redshift $z=3$.}
\label{plotz3}
\end{figure}

\begin{figure}
\centering
\epsfig{figure=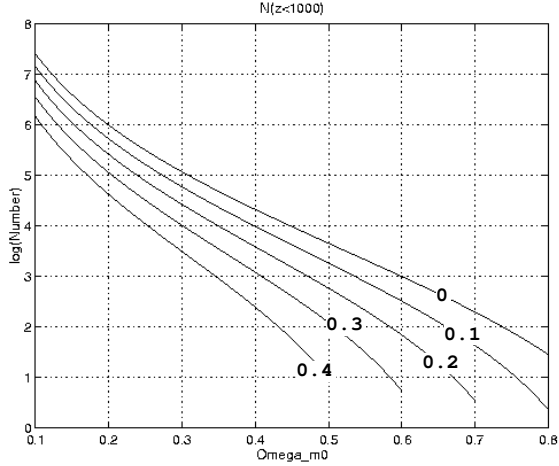, width=8cm}
\caption{Number of topological images within the observable universe (i.e.
with $z<1000$) in function of $\Omega_{m0}$
and $\Omega_{\Lambda0}$. This estimates the number of expected matched circles
on the cosmic microwave background.}
\label{plotz1000}
\end{figure}

As seen on figure \ref{plotz1}, detecting the topology with clusters
of galaxies would require both $\Omega_{\Lambda0}$ and $\Omega_{m0}$
to be very low. The situation is much better with groups of quasars
(figure \ref{plotz3}).  Figure \ref{plotz1000} shows the effect of the
two parameters ($\Omega_\Lambda$,$\Omega_m$) on the number of
topological images inside the observable universe.
This also provides an
estimation of the number of expected matched circles in the circle
method \cite{cornish97}.

Given standard values of the cosmological parameters \cite{moriond},
the numbers of pairs involving an object and one of its topological 
images will be
statistically low in compact hyperbolic universes. For instance a
cluster catalog will give no signature whatever the parameters and a
quasar catalog will typically require the cosmological constant to
vanish and the density parameter to be $\Omega_{m0}<0.4$. It can also
be seen that the generic effect of the cosmological constant is to make
the horizon volume bigger and thus to dilute the number of 
topological images.

\subsection{compact elliptic universes}\label{CEM}

We proceed as in the previous section but now the spatial sections are
of the form $S^3/\Gamma$, where the holonomy group $\Gamma$ is either a
cyclic group, a dihedral group or the symmetry groups $T,O,I$
respectively of the tetrahedron, octahedron or isocahedron (see
\cite{luminet95} for a complete description). The local geometry of
the background spacetime is described by a Friedmann-Lema\^\i tre
universe with the metric
\begin{equation}
ds^2=-dt^2+a^2(t)\left(d\chi^2+\sin^2{\chi}d\Omega^2 \right).\label{METR2}
\end{equation}

In units of the curvature radius (i.e. when $K=+1$), the volume of the
spatial sections is given by
\begin{equation}
\hbox{Vol}(S^3/\Gamma)=\frac{\hbox{Vol}(S^3)}{|\Gamma|}=\frac{2\pi^2}
{|\Gamma|},\label{vol_ell}
\end{equation}
where $|\Gamma|$ is the order of the group $\Gamma$
(e.g. $|\Gamma|=12,24,60$ respectively for $T,O,I$)
\cite{luminet95,wolf67}. Since the volume of the sphere of radius
$\chi$ is given by
\begin{equation}
\hbox{Vol}(\chi)=\frac{2\chi-\sin{2\chi}}{2\pi},
\end{equation}
the number of topological images defined as in \S\ref{CHM}, is 
\begin{equation}
N(\Omega_{m0},\Omega_{\Lambda0};z<Z)=\frac{|\Gamma|}{2\pi}
\left(2\chi(Z)-\sin{2\chi(Z)}\right).
\end{equation}
$\chi(Z)$ is computed as in equation (\ref{fried2}-\ref{chi1}) by
changing the sign of the curvature index $K$ in (\ref{fried1})
[note that it does not affect equation (\ref{fried2})]
so that
\begin{equation}
\chi(z)=
\int_{\frac{1}{1+z}}^1\frac{\sqrt{\Omega_{m0}+\Omega_{\Lambda0}-1}dx}
{x\sqrt{\Omega_{\Lambda0}x^2+(1-\Omega_{m0}-\Omega_{\Lambda0})
+\frac{\Omega_{m0}}{x}}}.\label{chi2}
\end{equation}
When $\Omega_{\Lambda}=0$, this expression can be computed analytically,
as in equation (\ref{chi_de_z}) \cite{peebles93}

\begin{figure}
\centering
\epsfig{figure=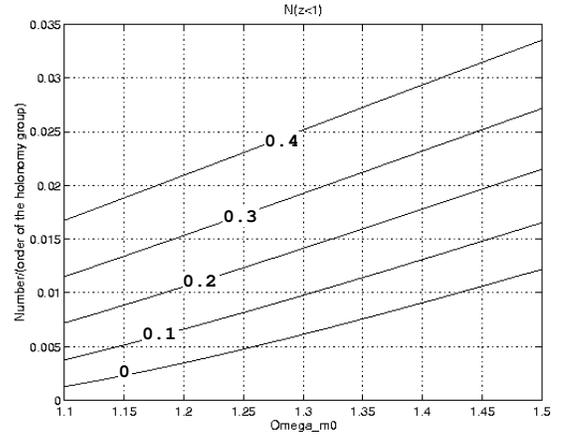, width=8cm}
\caption{Number of topological images in a galaxy cluster catalog
up to a redshift $z=1$ in function of $\Omega_{m0}$
and $\Omega_{\Lambda0}$ in a locally elliptic space.}
\label{zs3}
\end{figure}

\begin{figure}
\centering
\epsfig{figure=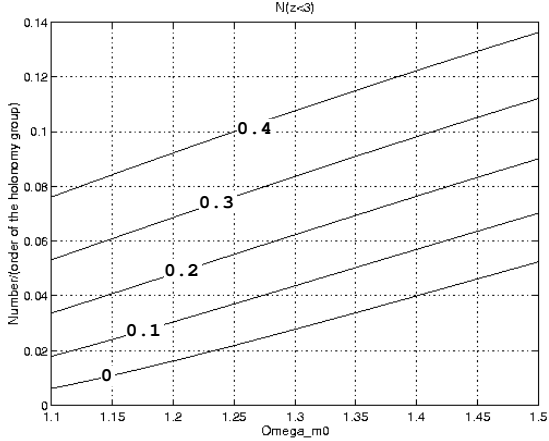, width=8cm}
\caption{Same as figure 4 for a catalog of quasars
up to a redshift $z=3$.}
\label{zs33}
\end{figure}

We plot $N/|\Gamma|$ in term of $\Omega_{m0}$ for different values of
$\Omega_{\Lambda0}$. It can be concluded that the topology of an
elliptic space can be detected respectively by a catalog of galaxy clusters
(see figure \ref{zs3})
only if $|\Gamma|>200$ and if $|\Gamma|>50$ in a catalog of quasars
(see figure \ref{zs33}) when $\Omega_\Lambda=0$.

A non vanishing cosmological constant improves the situation and
holonomy groups of lower order can be considered. Nevertheless, we are still
constrained by the fact that the total energy density has to be
compatible with the estimated age of the universe.

\section{Numerical results}\label{NUM}

In our numerical simulations, we concentrate only on compact 
hyperbolic models. 
We first generate an idealised catalog, ${\cal C}$, by distributing {\it
homogeneously} objects in the fundamental domain. A homogeneous
distribution is defined by the requirement that the number of objects
per unit volume is constant, i.e. by
\begin{equation}
\frac{dN}{dV}=constant,
\end{equation}
$dV$ being given by
\begin{eqnarray}
dV&=&\sinh^2{\chi}d\chi\sin\theta d\theta d\varphi\nonumber\\
&=&d[\cosh{2\chi}-2\chi]d\cos{\theta} d\varphi,
\end{eqnarray}
with $\cos{\theta}\in[-1,1]$ and $\varphi\in[-\pi,\pi]$.  We first
create a catalog in the smallest sphere containing the
fundamental polyhedron. Then, in order to obtain 
a set ${\cal C}_1$, we reject the points lying outside the
fundamental domain by checking if they are on the
same side of the faces of the fundamental domain than its
center. This can easily be achieved when we know the Minkowskian
coordinates of the vertices of the polyhedron, which can be obtained
from {\it SnapPea} (see Appendix A).

We then unfold the catalog by applying the
generators of the holonomy group to obtain the set ${\cal C}_2$
\begin{eqnarray}
{\cal C}_2=\left\lbrace\tilde x,\forall x\in{\cal C}_1,\quad
\tilde x=\sum_{g\in\Gamma}g(x);\quad
\chi(\tilde x)\leq\chi_{m}\right\rbrace,
\end{eqnarray}
where $\chi_{m}$ is the maximal value of the radial coordinate
of the set ${\cal C}_2$. To generate a catalog with a given depth
$z$ in redshift, we truncate ${\cal C}_2$ so that
\begin{equation}
{\cal C}(z)=\left\lbrace x\in{\cal C}_2;\quad 
\hbox{dist}[0,x]\leq\chi(\Omega_{m0},
\Omega_{\Lambda0};z)\right\rbrace,\label{select}
\end{equation}
where $\chi(\Omega_{m0},\Omega_{\Lambda0};z)$ is given by equation
(\ref{chi1}). This accounts as selecting the objects located within the
geodesic ball of radius $\chi(\Omega_{m0},\Omega_{\Lambda0};z)$
centered onto an observer placed at the centre.
 
We then compute all the three dimensional separations between all
the pairs of ${\cal C}(z)$ and plot the histogram of the
number of pairs with a given separation.\\

Indeed the former procedure applies when the observer stands at the
center of the polyhedron $(\chi=0)$. Now, if the observer is at
a position, $(\chi_\oplus\not=0,\theta_\oplus=0,\varphi_\oplus=0)$ say,
we have to perform a coordinate change to ``center''
the catalog on the observer before selecting the object as in
(\ref{select}). The Minkowskian coordinates, $x'$ say, of a point
in the frame centered on the observer are related to the ``old'' coordinates
$x$ (i.e in the frame centered on $\chi=0$) by
\begin{equation}
x'={\cal M}_{0\rightarrow\oplus}x\quad{\rm with}\quad
{\rm det}({\cal M}_{0\rightarrow\oplus})=1,
\end{equation}
where ${\cal M}_{0\rightarrow\oplus}$ is a matrix determined by the fact that
the image of the ``old'' center $\chi=0$ is the observer's position
$x_\oplus=(\cosh{\chi_\oplus}\equiv\gamma,0,0,
\sinh{\chi_\oplus}\equiv\beta\gamma)$ so that
\begin{equation}
{\cal M}_{0\rightarrow\oplus}=
\left(
\begin{array}{cccc}
\gamma&0&0&-\beta\gamma\\
0&1&0&0\\
0&0&1&0\\
-\beta\gamma&0&0&\gamma
\end{array}
\right).
\end{equation}
We recognize a Lorentz transformation. The same method applies
when $\theta_\oplus\not=0$ and $\varphi_\oplus\not=0$ but the matrices
are not so straightforward, so that we do not consider them here.

The catalog ${\cal C}(z)$ is then constructed as in (\ref{select}) but
using the points ${x'}$ instead of $x$, and the procedure
of pair computation is not affected.\\

\begin{figure}
\centering
\epsfig{figure=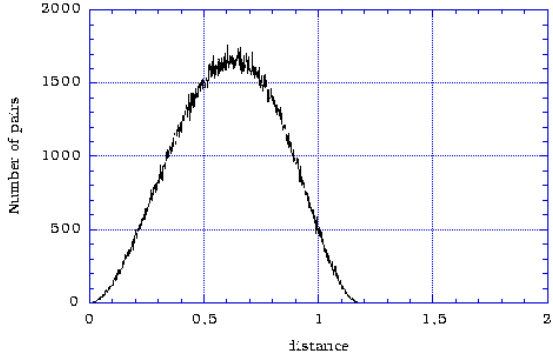, width=8cm}
\caption{Pair histogram for a galaxy cluster catalog ($z=1$) measured by an
 observer centered in $\chi=0$ in a hyperbolic universe whose spatial
 sections have the topology of the Weeks manifold and $\Omega_{m0} =
 0.2$; $\Omega_\Lambda=0$.}
\label{llu1}
\end{figure}

\begin{figure}
\centering
\epsfig{figure=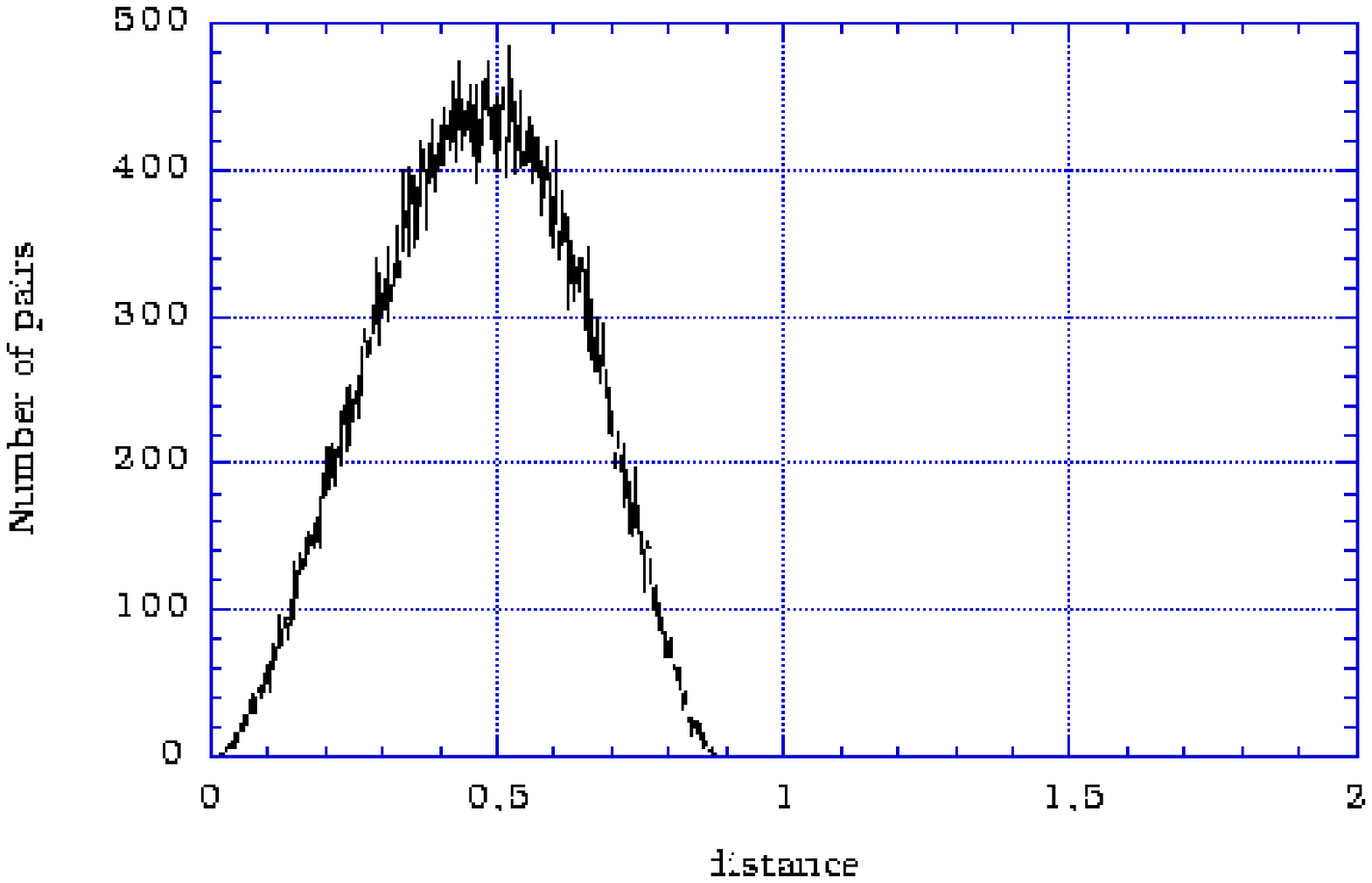, width=8cm}
\caption{Pair histogram for a galaxy cluster catalog ($z=1$) measured
 by an observer centered in $\chi=0$ in an universe whose spatial sections
 have the topology of the Weeks manifold and $\Omega_{m0}=0.5$.}
\label{llu3}
\end{figure}

\begin{figure}
\centering
\epsfig{figure=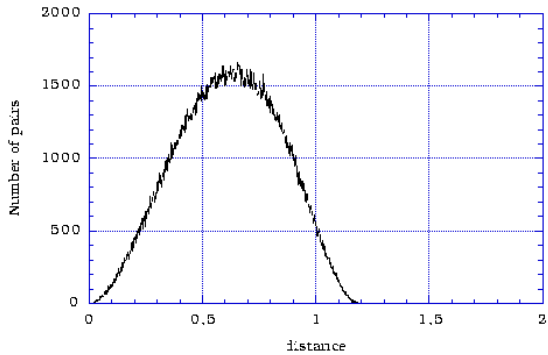, width=8cm}
\caption{Pair histogram for a galaxy cluster catalog ($z=1$) measured
 by an observer located near a face of the Weeks fundamental domain. Its Klein coordinates are
$x=(0,0,\tanh{0.5})$ and $\Omega_{m0}=0.2$.}
\label{llu2}
\end{figure}

\begin{figure}
\centering
\epsfig{figure=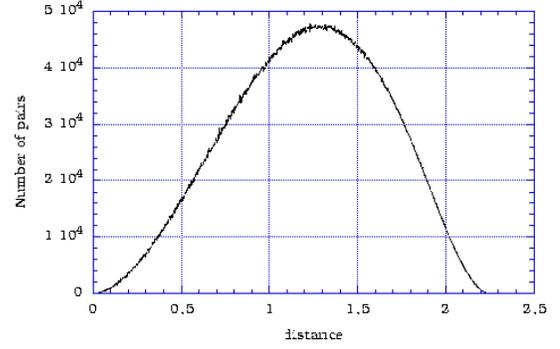, width=8cm}
\caption{Pair histogram for a quasar catalog ($z=3$) measured
 by an observer centered in $\chi=0$ in an universe whose spatial sections
 have the topology of the Weeks manifold and $\Omega_{m0}=0.2$.}
\label{llu4}
\end{figure}

We now generate some pair histograms in a universe whose spatial
sections have the topology of the Weeks manifold using $N=1000$ objects
in the fundamental domain. In figure (\ref{llu1}) we assume that
$\Omega_\Lambda=0$, $\Omega_{m0}=0.2$, that the observer stands at the
center of the polyhedron ($\chi=0$) and that he uses a galaxy cluster catalog
of depth $z=1$. We then study the dependence on $\Omega_{m0}$
(see figure \ref{llu3} where $\Omega_{m0}=0.5$), on the
position of the observer (see figure \ref{llu2} where the
observer stands near a face) and on the catalog depth
(see figure \ref{llu4} where we assume that the observer is using
a quasar catalog of depth $z=3$).

\begin{figure}
\centering
\epsfig{figure=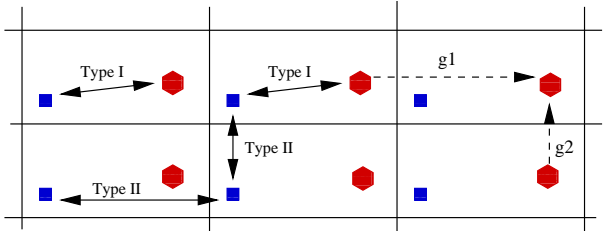, width=8cm}
\caption{The difference between type I and type II pairs on the
example of the two dimensional torus (the translations $g_1$, $g_2$
$g_1^{-1}$ and $g_2^{-1}$ are the generators of its holonomy group).
Type I pairs are the ones between the ghosts of two distinct objects
and type II pairs are the ones between two topological images of the
same object (hexagon or square).}
\label{pairtype}
\end{figure}

As a matter of fact, we do not observe any peaks in these histograms,
contrarily to the case of locally Euclidean universes.  Let us try to
understand why. Two kinds of pairs can give birth to peaks (see figure
\ref{pairtype}):
\begin{enumerate}
\item {\it Type I pairs} of the form $\lbrace g(x),g(y)\rbrace$, since
$\hbox{dist}[g(x),g(y)]=\hbox{dist}[x,y]$ for all points and all elements $g$ of
$\Gamma$.
\item {\it Type II pairs} of the form $\lbrace x,g(x)\rbrace$ if
$\hbox{dist}[x,g(x)]=\hbox{dist}[y,g(y)]$ for at least some points and elements $g$ of
$\Gamma$.
\end{enumerate}

Type I pairs are always present, whatever the topology. Their number 
roughly equals the number of copies of the fundamental domain 
within the catalog's limits.
Type II pairs produce peaks when the separation distance
between topological images is independent of the location of the source.

In compact locally Euclidean universes, type I and type II pairs are
both present.  The reason is that the 3-torus has the very special
property that the separation distance of gg-pairs (i.e. any pair of
images comprising an original and one of its ghosts, or two ghosts of
the same object) is independent of the location of the source.  In
other Euclidean spaces the spectrum of gg-pair distances varies with
the location of the source. However all closed Euclidean 3-manifolds
have the 3-torus as a covering space, so for each such manifold there
will be some distances which are independent of the location of the
source.  As a consequence, the topological signal expected in the
histogram from type I and type II pairs clearly stands out, as was
shown in the simulations of \cite{lehoucq96}. \\

\begin{figure}
\centering
\epsfig{figure=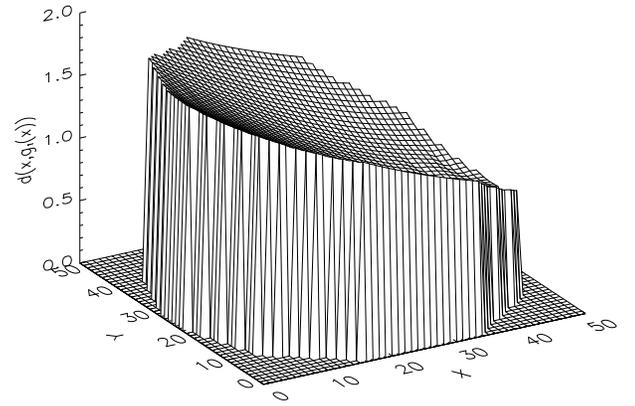, width=10cm}
\caption{The function $\hbox{dist}[x,g_1(x)]$ for all the points located within
the fundamental domain of Weeks manifold and on the disk ${\varphi=0}$
for the generator $g_1$ given in Appendix A. This illustrates the fact
that this function depends on $x$ in a locally hyperbolic manifold,
contrary to Euclidean manifolds.}
\label{llu5}
\end{figure}

In compact hyperbolic manifolds, $\hbox{dist}[x,g(x)]$ always depends on the
position $x$ of the source \cite{thurston}. This fact is clearly
illustrated by the numerical calculation of figure \ref{llu5}.  Thus
type II pairs cannot appear (see figure \ref{pairtype}). Moreover, as
shown in \S \ref{CHM}, the number of type I pairs is too low to
generate significant peaks in the distance histogram.  Hence the
crystallographic method fails.

The small number of Type I pairs in hyperbolic manifolds is due to the
property that the volume of the manifold is fixed once the topology is
determined (the {\it rigidity theorem}) contrary to Euclidean spaces
where the characteristic sizes and the volume of the fundamental
polyhedron can be choosen at will (since $K=0$ the geometry does not
impose any characteristic size).\\

In elliptic spaces, distances are position-independent whenever the
holonomy is a Clifford translation \cite{weeksprivate}.  A Clifford
translation is an isometry $g$ such that the displacement function
$\hbox{dist}(x,g(x))$ is constant. This is precisely what is required to get
type II pairs in the histogram. All finite groups of Clifford
translations of spheres are the cyclic group, the binary dihedral,
tetrahedral, octahedral and icosahedral groups \cite{wolf67}.  Next
(theorem 7.6.7 in \cite{wolf67}): $S^3/\Gamma$ is a Riemannian
homogeneous elliptic space if and only if $\Gamma$ is a group of
Clifford translations of $S^3$.  Given the classification of
three--dimensional spherical space forms (see \cite{luminet95}, \S 7),
we deduce that all homogeneous elliptic spaces usable for cosmology,
such as lens spaces $L(p) =S^3/Z_{p}$ or the Poincar\'e dodecahedral
space, satisfy this property. The covering transformations which take
a source to its nearest neighbours are Clifford translations (although
the transformations to more distant neighbours might not be), and type
II pairs can be produced.\\

\section{conclusion}

In this article, we generalised the crystallographic method
based on the existence of topological images, to universes
with non-Euclidean compact spatial sections.

The analysis was performed by using the smallest known compact
hyperbolic manifold, where we expect to have the greatest number of
topological images. It turns out that we do not observe, contrary to
Euclidean universes, any peaks in the pair 3D separation
histogram.

The absence of peaks is due to combined effects,
of the mathematics and of the cosmological parameters.
\begin{enumerate}
\item Locally hyperbolic manifolds are such that
$\hbox{dist}[x,g_n(x)]$ depends on $x$, so that
there is no amplification for the {\it type II pairs} $\lbrace x,g_n(x)
\rbrace$,
whereas $\hbox{dist}[x,g_n(x)]=\hbox{dist}[y,g_n(y)]$ in the Euclidean case.
This suppresses the peaks.
\item The peaks associated to the isometries (i.e. such that $\forall
g\in\Gamma,\quad \hbox{dist}[g(x),g(y)]=\hbox{dist}[x,y]$) must remain. But,
given the cosmological parameters, we have shown in \S \ref{CHM} that
the number of topological images is too low to create such peaks
associated to {\it type I pairs}.
\end{enumerate}

In elliptic universes, we have studied the influence of the cosmological
parameters. As in the hyperbolic case, type I pairs can be observed for
very small universes only and, as in the Euclidean case, type II pairs
may be present, due to the fact that the holonomies are Clifford translations.
However, such universes are not favored by the present estimates
of the cosmological parameters \cite{moriond}.

We conclude that in practice, the crystallographic method will be able
to detect the topology only if the universe is locally Euclidean. Such
universes have the interesting property that the characteristic sizes
of their fundamental domain are decoupled from the cosmological
parameters and thus from the Hubble radius.  Whatever the underlying
geometry, discrete sources such as quasars, X-ray galaxy clusters or
infrared galaxies can still help to investigate the cosmic topology by
looking for multiple images of individual objects.

\section*{Acknolewdgments}
We are grateful to J. Weeks and H. Fagundes for their precious comments.


\onecolumn
\pagebreak

\onecolumn
\section*{Appendix A~: Description
of the Weeks manifold}

\begin{figure}
\centering
\epsfig{figure=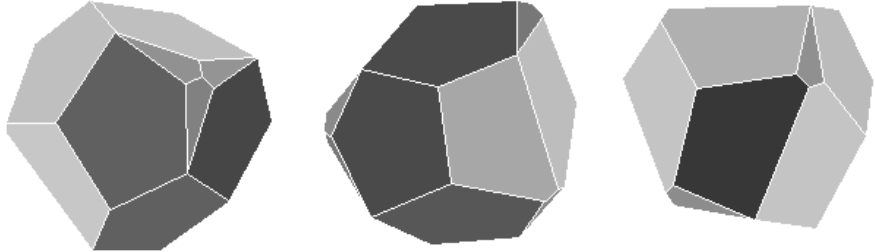, width=12cm}
\caption{Three views of the fundamental domain of the Weeks manifold.}
\label{weeks_FP}
\end{figure}

We considered the Weeks manifold [closed census  m003(-3,1)]. Its
fundamental polyhedron has 18 faces and 26 vertices (see figure
\ref{weeks_FP}). All the following quantities are needed
to perform our computation and can be obtained from the software
{\it SnapPea}. The volume of the manifold, in units of the curvature
radius, is $0.94272$.

The Klein coordinates (\ref{Kleincoord}) of the 26 vertices are

\begin{center}
\begin{tabular}{|r|rrr|}
\hline
label&$X_1\quad\,$&$X_2\quad\,$&$X_3\quad\,$\\
\hline
0&  0.10797407&  0.34689848& -0.41772745\\
1& -0.00056561& -0.36314169& -0.51501423\\
2&  0.03670097& -0.29313316& -0.54565764\\
3& -0.08402116& -0.35849225& -0.49943971\\
4& -0.23493634&  0.01568564& -0.59147636\\
5& -0.08019087&  0.60971881 & 0.08263574\\
6&  0.00049690&  0.31902895&  0.45245272\\
7& -0.42135580& -0.01132323&  0.46844736\\
8& -0.44370265&  0.45474638& -0.04023802\\
9& -0.03061224 & 0.62921121&  0.01637099\\
10& -0.06244403&  0.61432068& -0.06105007\\
11& 0.52204774&  0.13760656& -0.30585432\\
12& -0.51128589&  0.15380739& -0.31614284\\
13& -0.50817234& -0.21892234& -0.01808485\\
14&  0.04566464& -0.41129238& -0.46235223\\
15& -0.18945313& -0.58387452& -0.16877044\\
16& -0.34964269&  0.00034506&  0.51260720\\
17& -0.43363513& -0.08848249&  0.43491114\\
18& -0.11977854& -0.34363564&  0.52235338\\
19& -0.01122014& -0.58564972&  0.20470658\\
20&  0.56409719&  0.28438251&  0.07878044\\
21&  0.43096130&  0.11398724&  0.43161967\\
22&  0.45298033&  0.03203520&  0.43691467\\
23&  0.49689789& -0.00018511&  0.37162982\\
24&  0.39981755& -0.37241380&  0.08914633\\
25&  0.42311163& -0.25223127& -0.40328816\\
\hline
\end{tabular}
\end{center}

The 18 faces are then defined by their vertices as

\begin{center}
\begin{tabular}{|c|c|ccccc|}
\hline
faces&Number of&label of &label of&label of&label of&label of\\
&vertices&vertex 1&vertex 2&vertex 3&vertex 4&vertex 5\\
\hline
I&5&  23&  24 &  25&  11&  20\\
II&5&   2 &  4 &  0&  11&  25\\
III&5&  22 & 18 & 19&  24 & 23\\
IV&5&  10 &  0&   4&  12 &  8\\
V&5&   8  & 7&  16 &  6&   5\\
VI& 5&  15 & 13&  12 &  4 &  3\\
VII&5&  17 & 13 & 15&  19 & 18\\
VIII&5&  14 & 25 & 24 & 19&  15\\
IX&5&  18 & 22 & 21 &  6&  16\\
X&5&   9&  20 & 11&   0 & 10\\
XI&5&   7 &  8 & 12&  13 & 17\\
XII&5&   5 &  6&  21&  20 &  9\\
XIII&4&   3 &  4 &  2&   1&\\
XIV&4&  20&  21&  22&  23&\\
XV&4&   7&  17 & 18&  16&\\
XVI&4&   1&   2&   25&  14&\\
XVII&4&   1&  14 & 15&   3&\\
XVIII&4&  10&   8 &  5&   9&\\
\hline
\end{tabular}
\end{center}

The 18 generators of the holonomy group can then be written as

$$
g_1=\left(
\begin{array}{rrrr}
1.58926252069783 &  -0.40490373463745&   -0.18520828837947&    1.15217455938197\\
1.19813100468664 &  -0.37867518857251&   -0.43845341527127&    1.44909682512520\\
-0.29994054013040 &   0.58749594014885&   -0.79530849910863&   -0.33510780230833\\
-0.01652657525663 &   0.82182763167926&    0.45776096815586&    0.33959883323000
\end{array}\right)
$$
$$
g_2=\left(
\begin{array}{rrrr}
    1.58926252069784  &   0.58852165407720 &   -0.78772550995989 &   -0.74758688012532\\
    0.60542090101752 &    0.63256516975419 &   -0.95802870162689 &    0.22040140712305\\
   -0.74661392026174 &   -0.95121865744570 &    0.02947692798603 &    0.80730819540762\\
    0.77575031187707 &   -0.20347508215704 &   -0.83774912510678&   -0.92658666119331
\end{array}\right)
$$
$$
g_3=\left(
\begin{array}{rrrr}
    1.58926252069784 &   0.90966168251236&    0.52649891309985&    0.64889897331428\\
   -0.39527310953786 &  -0.96583883836601 &  -0.09504624064542&    0.46299285260867\\
    0.83783613378036 &   0.46865953388686 &   1.11407913829435&    0.49107565810403\\
   -0.81703435760262 &  -0.82158259245920 &   0.16430152540933&   -0.98262515437950
\end{array}\right)
$$

$$
g_4=\left(
\begin{array}{rrrr}
    1.58926252069784&   -0.50251081201258&    1.10144921603985&    0.24504666492427\\
   -0.80335677967765&    0.05959497029949&   -0.93768098628153&   -0.87326108520599\\
   -0.85819691489983&    0.41642334058451&   -1.15485297933977&    0.47896570051526\\
    0.37930370351109&   -1.03709071788392&    0.01610402406107&    0.26087315458913
\end{array}\right)
$$

$$
g_5=\left(
\begin{array}{rrrr}
    1.58926252069784&   -0.11125231907787&   -0.10906253703873&   -1.22535041690502\\
    0.61314085066332&   -0.72730066360826&   -0.62733910551353&   -0.67336549819589\\
    0.97779233966538&    0.49501806133898&   -0.13277443919182&   -1.30130931248255\\
   -0.44015428821540&    0.48822939502184&   -0.77505563153628&    0.59553053934699
\end{array}\right)
$$
$$
g_6=\left(
\begin{array}{rrrr}
    1.58926252069784&   -0.48013536446388&   -0.94330313872012&   -0.63671389176518\\
   -1.21801166215147&    0.63453136165388&    1.27966469214643&    0.66586833224629\\
    0.12199830361655&    0.75940754853765&   -0.15821652454111&   -0.64276845961162\\
    0.16528570559051&   -0.50119865773022&    0.47670391078232&   -0.74085940056587
\end{array}\right)
$$
$$
g_7=\left(
\begin{array}{rrrr}
    2.07713761580105&   -1.55800356296413&   -0.47129628725962&   -0.81547862165852\\
   -0.34790844003080&    0.59158539050820&   -0.84366445697397&    0.24351035382656\\
   -0.55790510170459&    0.95617333032507&    0.27707820937007&   -0.56587837100861\\
   -1.69770500671437&    1.47075980034853&    0.65846650660136&    1.13379440781211
\end{array}\right)
$$
$$
g_8=\left(
\begin{array}{rrrr}
    2.07713761580106&   -0.62403517630796&    0.95362111998219&    1.41974910925485\\
   -0.85113707827869&    0.92242574522307&    0.12808289744563&   -0.92582927257265\\
   -0.52567197801091&   -0.52156865254264&    0.00566611122130&   -1.00213026315664\\
    1.52109674922046&   -0.51625263825403&    1.37584734147421&    1.07436615124600
\end{array}\right)
$$
$$
g_9=\left(
\begin{array}{rrrr}
    2.07713761580107&    0.70433686017330&   -1.67848312773007&    0.03323631000097\\
   -0.28072927025369&    0.14834866484720&    0.42816244471325&    0.93460072638127\\
   -1.44346161316703&   -0.14613011383240&    1.71843229305311&   -0.33048127437171\\
   -1.07336402171483&   -1.20529211233965&    0.82521087317360&     0.13567704979006
\end{array}\right)
$$

The nine other matrices are defined by $g_{k+9}=g_k^{-1}$
and any element $g\in\Gamma$ can be written as
$$g=\prod_{i\in I}g_{n_{i}};\qquad n_i\in\lbrace1,..,18\rbrace.$$


\end{document}